\documentclass[preprintnumbers,showpacs]{revtex4}
\usepackage{axodraw}
\usepackage{graphicx}
\begin{document}

\title{
Study of color suppressed modes $B^0 \to \bar D^{(*)0} \eta^{(\prime)} $  }

\author{
Cai-Dian L\"u \\
\small CCAST (World Laboratory), P.O. Box 8730,
Beijing 100080, China;\\
\small Institute of High Energy Physics, CAS, P.O. Box 918(4),
Beijing 100039,
China\thanks{Mailing address}}

  \preprint{BIHEP-TH-2003-23}

\begin{abstract}
 The color suppressed modes $B^0 \to \bar D^{(*)0}
\eta^{(\prime)} $ are analyzed in perturbative QCD approach.  We
find that the dominant contribution is from the non-factorizable
diagrams. The branching ratios calculated in our approach for $B^0
\to \bar D^{(*)0} \eta $ agree with current experiments. By
neglecting the gluonic contribution, we predict the branching
ratios of $B^0 \to \bar D^{(*)0} \eta'$ are at the comparable size
of $B^0 \to \bar D^{(*)0} \pi^0$, but smaller than that of $B^0
\to \bar D^{(*)0} \eta $.
\end{abstract}


 \maketitle

\section{INTRODUCTION}

     The hadronic decays   $B^0 \to \bar D^{(*)0} \eta^{(\prime)} $ are
     color suppressed modes, which belong to class II decays in
     the factorization approach (FA) \cite{neu}.
The relevant effective weak Hamiltonian for these decays is given by
\begin{eqnarray}
{\cal H}_{\rm eff} = {G_F\over\sqrt{2}}\, V_{cb}V_{ud}^*
\Big[C_1(\mu)O_1(\mu)+C_2(\mu)O_2(\mu)\Big]\;,
\end{eqnarray}
where the four-quark operators are
\begin{eqnarray}
O_1= (\bar db)_{V-A}(\bar cu)_{V-A}\;,\qquad\qquad
O_2= (\bar cb)_{V-A}(\bar du)_{V-A}\;,
\end{eqnarray}
with the definition $(\bar q_1q_2)_{V-A}\equiv \bar
q_1\gamma_\mu(1- \gamma_5)q_2$.  The Wilson coefficients $C_1 \sim
-0.2$ and $C_2 \sim 1$ are calculated at $m_b$ scale. The main
contribution of these decays in FA is proportional to the Wilson
coefficients $a_2 = C_1 +C_2/3$, which is a small number. That is
the reason why class II decays usually have small branching
ratios. Theoretical study of $B^0 \to \bar D^{(*)0}
\eta^{(\prime)} $ decays gives a branching ratio of $10^{-5}$
\cite{deta}.
 However, recent experiments by Belle and BABAR show
that the branching ratios of class II decays are not so small
\cite{BelleC,CLEOC}. The branching ratios of $B^0 \to \bar
D^{(*)0} \eta $ are of $10^{-4}$. Although the gluonic mechanism
can enhance the $B^0 \to \bar D^{0} \eta^{\prime} $ decay
branching ratio to $10^{-4}$ \cite{glu}, it may be difficult to
explain the large branching ratio of $B^0 \to \bar D^{(*)0} \eta
$. It means that the non-factorizable contributions in these
decays are very important. This is confirmed in the recent
theoretical study on charmed final state B meson decays in
perturbative QCD approach \cite{dpi}.

 The perturbative QCD approach (PQCD) for the exclusive
  hadronic B decays was developed some time ago \cite{LB,BS},
  and applied to the semi-leptonic \cite{TLS} and non-leptonic decays
  \cite{KLS,LUY,kou}
  successfully.    In this formalism, factorizable contributions,
  non-factorizable  and annihilation contributions are all
  calculable. By including the $k_T$ dependence of the wave
  functions and Sudakov form factor, this approach is free of
  endpoint singularity.
   Recent study shows that PQCD approach works well for
  charmless B decays \cite{KLS,LUY,kou}, as well as for channels
  with one charmed meson in the final states \cite{dpi,dphi}.
    We will show the PQCD calculation of $B^0 \to \bar D^{(*)0} \eta^{(\prime)} $
    decays in the next
  section, and discuss the numerical results in section \ref{s3}. The
  conclusion is in section~\ref{s4}.

\section{$B^0 \to \bar D^{(*)0} \eta^{(\prime)} $ DECAY AMPLITUDES IN PQCD}

In two-body hadronic B decays, the two outgoing mesons are
energetic. Each of the valence quarks inside these mesons  carries
large momentum. Most of the energy comes from the heavy b quark
decay in quark level. The light quark (d quark) inside $B^0$
meson, which is usually called spectator quark, carries small
momentum at order of $\Lambda_{QCD}$. This quark also goes into
final state meson in spectator diagrams. Therefore, we need an
energetic gluon to connect this quark to the four quark operator
involved in the b quark decay. Such that the spectator quark get
energy from the four quark operator to form a fast moving light
meson. The hard four quark dynamic together with the spectator
quark becomes six-quark effective interaction. Since six-quark
interaction is hard dynamics, it is perturbatively calculable. The
non-perturbative dynamics in this process is described by the wave
functions of mesons consisting of quark and anti-quark pair. The
decay amplitude is then expressed as
\begin{eqnarray}
 \mbox{Amplitude}
\sim \int\!\! d^4k_1 d^4k_2 d^4k_3\ \mathrm{Tr} \bigl[ C(t)
\Phi_B(k_1) \Phi_{D^{(*)}}(k_2) \Phi_{\eta^{(\prime)}}(k_3) H(k_1,k_2,k_3, t)
e^{-S(t)}\bigr]. \label{eq:convolution1}
\end{eqnarray}
Here $C(t)$ is the QCD corrected Wilson coefficient of the
relevant four quark operator at scale $t$. Although next-to
leading order results have been given \cite{buras}, we will use
leading order one here \cite{LUY}. $\Phi_i$ are the meson wave
functions, which include the non-perturbative contributions in
these decays. The non-perturbative wave functions are not
calculable in principal. But they are universal for all the
hadronic decays. We will use the ones determined from other
measured decay channels \cite{dpi,KLS,LUY,kou}. The exponential
$S(t)$ is the so-called Sudakov form factor, which includes the
double logarithm resulting from the resummation of soft and
collinear divergence.  This form factor is also calculated to
next-to leading order in the literature \cite{LY1}. The Sudakov
factor effectively suppresses the soft contributions in the
process \cite{dpi,KLS,LUY}, thus it makes the perturbative
calculation of hard part reliable.

 Now the only left part of the decay amplitude is the hard part
 $H(t)$. Since it involves the four quark operator and the spectator
 quark
 connected by a hard gluon, it is channel dependent, but perturbatively calculable. There
 are altogether 8 kinds of diagrams in our  $B^0 \to \bar D^{(*)0} \eta^{(\prime)} $
 decays, which are shown in Fig.1 for the spectator diagrams  and Fig.2 for the
 annihilation type
  diagrams.
 Notice that in Fig.1, the $\eta^{(\prime)}$ meson consists of $d
 \bar d$ content, while in Fig.2, it is a $u \bar u $ pair making
 $\eta^{(\prime)}$ meson. Since $\eta^{(\prime)}$ meson is isospin
 singlet ($u \bar u + d \bar d$), these two sets of diagrams give
 relatively positive contributions. On the other hand, in case of
 $B^0 \to \bar D^{(*)0} \pi^0$ decays \cite{dpi}, where $\pi^0$ is
 isospin triplet ($u \bar u - d \bar d$),
 these two sets of diagrams give destructive contributions there.
 Fortunately, as we will see later in the next section, the
 annihilation type diagrams are suppressed comparing to the
 spectator diagrams. Therefore, the branching ratios of these two
 kinds of decays are still comparable.

  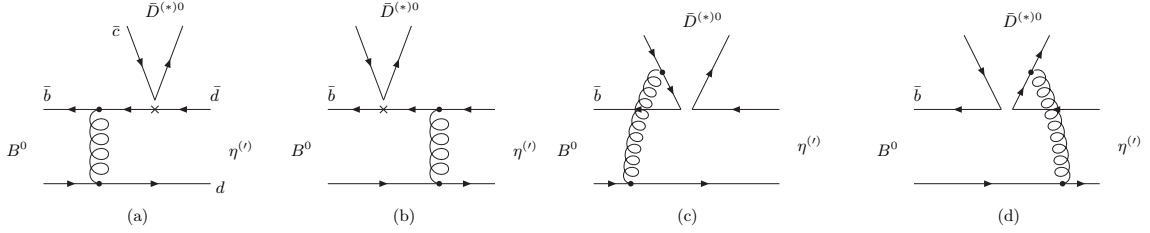
\begin{figure}
   \scalebox{0.7}{
   {
     \begin{picture}(140,120)(-30,0)
      \ArrowLine(30,60)(0,60)
      \ArrowLine(60,60)(30,60)
      \ArrowLine(90,60)(60,60)
      \ArrowLine(30,20)(90,20)
      \ArrowLine(0,20)(30,20)
      \Gluon(30,20)(30,60){5}{4} \Vertex(30,20){1.5} \Vertex(30,60){1.5}
      \Line(58,62)(62,58)
      \Line(58,58)(62,62)
      \ArrowLine(45,105)(60,65)
      \ArrowLine(60,65)(75,105)
      \put(-20,35){$B^0$}
      \put(100,35){$\eta^{(\prime)}$}
      \put(55,110){$ \bar D^{(*)0}$}
      \put(0,65){\small{$\bar{b}$}}
      \put(37,100){\small{$\bar{c}$}}
      \put(90,65){\small{$\bar{d}$}}
      \put(93,15){\small{${d}$}}
      \put(45,0){(a)}
   \end{picture}
   }}
   \scalebox{0.7}{
   {
     \begin{picture}(140,120)(-30,0)
        \ArrowLine(30,60)(0,60)
        \ArrowLine(60,60)(30,60)
        \ArrowLine(90,60)(60,60)
        \ArrowLine(60,20)(90,20)
        \ArrowLine(0,20)(60,20)
        \Gluon(60,20)(60,60){5}{4} \Vertex(60,20){1.5} \Vertex(60,60){1.5}
        \Line(28,62)(32,58)
        \Line(28,58)(32,62)
        \ArrowLine(15,105)(30,65)
        \ArrowLine(30,65)(45,105)
        \put(-20,35){$B^0$}
        \put(100,35){$\eta^{(\prime)}$}
        \put(28,110){$ \bar D^{(*)0}$}
        \put(15,48){\small{}}
        \put(35,66){\small{}}
        \put(0,65){\small{$\bar{b}$}}
        \put(35,0){(b)}
   \end{picture}
   }}
   \scalebox{0.7}{
   {
     \begin{picture}(170,110)(-20,0)
        \ArrowLine(47,60)(0,60)
        \ArrowLine(100,60)(53,60)
        \ArrowLine(20,20)(100,20)
        \ArrowLine(0,20)(20,20)
        \ArrowLine(37,80)(47,60)
        \ArrowLine(27,100)(37,80)
        \ArrowLine(53,60)(73,100)
        \Vertex(20,20){1.5} \Vertex(37,80){1.5}
        \GlueArc(150,20)(130,152,180){4}{9}
        \put(-20,35){$B^0$}
        \put(110,38){$\eta^{(\prime)}$}
        \put(48,105){$ \bar D^{(*)0}$}
        \put(15,48){\small{}}
        \put(55,48){\small{}}
        \put(0,65){\small{$\bar{b}$}}
        \put(45,0){(c)}
     \end{picture}
     \begin{picture}(140,110)(-20,0)
        \ArrowLine(47,60)(0,60)
        \ArrowLine(100,60)(53,60)
        \ArrowLine(80,20)(100,20)
        \ArrowLine(0,20)(80,20)
        \ArrowLine(27,100)(47,60)
        \ArrowLine(63,80)(73,100)
        \ArrowLine(53,60)(63,80)
        \Vertex(80,20){1.5} \Vertex(63,80){1.5}
        \GlueArc(-50,20)(130,0,28){4}{9}
        \put(-20,35){$B^0$}
        \put(110,38){$\eta^{(\prime)}$}
        \put(48,105){$\bar D^{(*)0}$}
        \put(15,48){\small{}}
        \put(55,48){\small{}}
        \put(0,65){\small{$\bar{b}$}}
        \put(45,0){(d)}
     \end{picture}
   }}
   \caption{Color-suppressed emission diagrams contributing to the
   $B^0\to \bar D^{(*)0} \eta^{(\prime)} $ decays.}
  \end{figure}

 The structures of the meson wave functions are
 \begin{eqnarray}
  B_{in}(P): [\not P + m_B ]\gamma_5 \phi_B(x)\;,\\
  D_{out}(P):  \gamma_5[\not P + m_D ]\phi_D(x)\;,\\
   D^{*}_{out}(P):  \not \epsilon[\not P + m_{D^*} ]\phi_{D^*}(x)\;,\\
  \eta^{(\prime)}_{out}(P):
 \gamma_5 [\not P \phi_A(x) + m_0 \phi_P(x)
      +\zeta m_0(\not n_-\not n_+ -1)\phi_T(x)]\;,
 \end{eqnarray}
 with $m_0 \equiv m_{\pi}^2/(m_u + m_d)=1.4$ GeV, utilizing
 isospin symmetry.
 And the light-like vectors are defined as $n_+=(1,0,{\bf 0}_T)$ and
 $n_- = (0,1,{\bf 0}_T)$. As shown in ref.\cite{TLS}, $\phi_B$ is identified as
 $\phi_+$ and the contribution of another $B$ meson wave function $\bar\phi_B\propto
 \phi_+-\phi_-$ is smaller in the PQCD calculations, therefore we neglect it.
Applying for heavy quark symmetry, there is only one independent
distribution amplitude $\phi_{D^{(*)}}$ in the heavy $D^{(*)}$
meson wave function \cite{dpi,dphi}. However, there are three
distribution amplitudes for the light $\eta^{(\prime)}$ meson wave
functions \cite{kou}, like the $\pi$ meson wave function. The
coefficients $\zeta = +1$ are for
 $\eta^{(\prime)}_{out}$ with $\bar u$ ($\bar d$) carrying the momentum
 $x_3P_3$, while $\zeta = -1$ for $\eta^{(\prime)}_{out}$ with $u$ ($d$) carrying the
 momentum $x_3P_3$.

 The gluonic mechanism  of $\eta'$ may make sizable
 contributions in   the  $B\to D^{(*)} \eta^{\prime} $ decays, but
 they are usually model dependent \cite{glu}.
 Thus we will not consider it here.
 The $\eta$ and $\eta'$ mesons are mixtures of  flavor SU(3) octet ($\eta_8$) and singlet
 ($\eta_0$) states in a two-mixing-angle formalism \cite{mix},
   \begin{eqnarray}
   \eta=  \cos \theta_8 |\eta_8\rangle-\sin \theta_0 |\eta_0 \rangle\;,
          \nonumber   \\
   \eta '=  \sin \theta_8|\eta_8\rangle+\cos \theta_0|\eta_0
   \rangle\;.\label{mic}
    \end{eqnarray}

 The definitions of the decay constants of $\eta$ and $\eta'$ are as follows:
 \begin{equation}
 \langle 0| \bar u \gamma_\mu \gamma_5 u | \eta^{(\prime)}
 (p)\rangle = if^u_{\eta^{(\prime)}} p_\mu ,~~~~~~~~~~
 \langle 0| \bar d \gamma_\mu \gamma_5 d | \eta^{(\prime)}
 (p)\rangle = if^d_{\eta^{(\prime)}} p_\mu .
 \end{equation}
 The $\bar ss$ components of $\eta$ and $\eta'$ are not relevant
 in our decay channels. Therefore we did not show them.
 The decay constants in the two-angle mixing formalism are
  \begin{eqnarray}
   f^u_{\eta} =f^d_{\eta} = \frac{f_8}{\sqrt{6}} \cos \theta_8 -\frac{f_0}{\sqrt{3}}
   \sin \theta_0 ,   \\
   f^u_{\eta^{\prime}}=  f^d_{\eta^{\prime}}=  \frac{f_8}{\sqrt{6}} \sin \theta_8 +\frac{f_0}{\sqrt{3}}
   \cos \theta_0 .
  \end{eqnarray}
The parameters are determined to be \cite{mix}
\begin{equation}
\begin{array}{ll}
\theta_8=  -22^\circ \sim -21^\circ ,     & f_8= 1.28f_\pi , \\
\theta_0=  -9^\circ \sim -4^\circ    ,  & f_0= (1.20-1.25)f_\pi \
.
\end{array}\label{ran}
\end{equation}

  The $\eta$ and $\eta'$ meson can also be expressed as a mixing
  in the quark flavor basis \cite{fit}
\begin{equation}
\left( \begin{array}{c} \eta \\ \eta'\end{array} \right) = \left (
\begin{array}{cc}
\cos \alpha & -\sin \alpha \\
\sin \alpha &   \cos \alpha
\end{array} \right) \left(
\begin{array}{c}
(u \bar u+d \bar d) /\sqrt{2} \\
s \bar s
\end{array} \right), \label{def}
\end{equation}
where $\alpha = \pi +\theta -\arctan (1/\sqrt{2})$ describe the
deviation from ideal mixing. The angle $\theta$ is one-angle
mixing parameter. From the above eqn. (\ref{def}), applying
isospin symmetry, we have
  \begin{eqnarray}
   f^u_{\eta} =f^d_{\eta} = f_\pi \cos \alpha/\sqrt{2} ,   \\
   f^u_{\eta^{\prime}}=  f^d_{\eta^{\prime}}= f_\pi \sin \alpha/\sqrt{2}
   .
  \end{eqnarray}
 The range of mixing parameters are determined to be  $\theta =-17^\circ \sim
-11^\circ$ \cite{fit}.

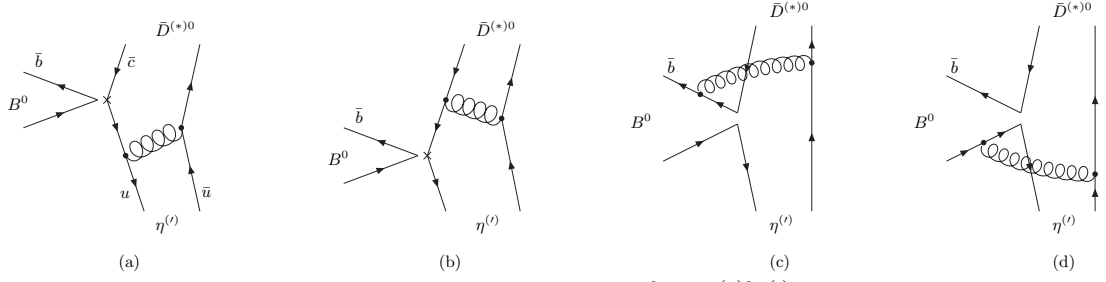
\begin{figure}
 \scalebox{0.7}{
 {
   \begin{picture}(130,120)(0,-20)
      \put(45,0){
      \rotatebox{90}{
        \ArrowLine(30,50)(0,40)
        \ArrowLine(60,60)(30,50)
        \ArrowLine(90,50)(60,60)
        \ArrowLine(45,20)(90,10)
        \ArrowLine(0,10)(45,20)
        \Gluon(45,20)(30,50){5}{4} \Vertex(45,20){1.5} \Vertex(30,50){1.5}
        \Line(58,62)(62,58)
        \Line(58,58)(62,62)
        \ArrowLine(45,105)(60,65)
        \ArrowLine(60,65)(75,105)
        \put(65,66){\small{}}
      }}
      \put(0,55){$B^0$}
      \put(80,95){$ \bar D^{(*)0}$}
      \put(80,-10){$\eta^{(\prime)}$}
      \put(15,78){\small{$\bar{b}$}}
        \put(65,78){\small{$\bar{c}$}}
     \put(105,8){\small{$\bar{u}$}}
      \put(61,7){\small{${u}$}}
       \put(65,55){\small{}}
      \put(60,-30){(a)}
   \end{picture}
   \begin{picture}(110,120)(0,-20)
      \put(85,0){
      \rotatebox{90}{
        \ArrowLine(30,60)(0,50)
        \ArrowLine(60,50)(30,60)
        \ArrowLine(90,40)(60,50)
        \ArrowLine(50,20)(90,10)
        \ArrowLine(0,10)(50,20)
        \Gluon(50,20)(60,50){5}{4} \Vertex(50,20){1.5} \Vertex(60,50){1.5}
        \Line(28,62)(32,58)
        \Line(28,58)(32,62)
        \ArrowLine(15,105)(30,65)
        \ArrowLine(30,65)(45,105)
        \put(35,66){\small{}}
      }}
      \put(40,25){$B^0$}
      \put(120,95){$\bar D^{(*)0}$}
      \put(120,-10){$\eta^{(\prime)}$}
      \put(55,48){\small{$\bar{b}$}}
      \put(105,25){\small{}}
       \put(100,-30){(b)}
         \end{picture}
 }}
 \scalebox{0.7}{
 {
   \begin{picture}(110,130)(0,-20)
      \put(195,0){
      \rotatebox{90}{
        \ArrowLine(47,60)(0,50)
        \ArrowLine(100,50)(53,60)
        \ArrowLine(80,20)(100,20)
        \ArrowLine(0,20)(80,20)
        \ArrowLine(27,100)(47,60)
        \ArrowLine(63,80)(73,100)
        \ArrowLine(53,60)(63,80)
        \Vertex(80,20){1.5} \Vertex(63,80){1.5}
        \GlueArc(-50,20)(130,0,28){4}{9}
      }}
      \put(80,45){$B^0$}
      \put(155,-10){$\eta^{(\prime)}$}
      \put(155,105){$\bar D^{(*)0}$}
      \put(140,37){\small{}}
      \put(140,55){\small{}}
      \put(100,75){\small{$\bar{b}$}}
      \put(155,-30){(c)}
         \end{picture}
   \begin{picture}(140,130)(0,-20)
      \put(235,0){
      \rotatebox{90}{
        \ArrowLine(47,60)(0,50)
        \ArrowLine(100,50)(53,60)
        \ArrowLine(20,20)(100,20)
        \ArrowLine(0,20)(20,20)
        \ArrowLine(37,80)(47,60)
        \ArrowLine(27,100)(37,80)
        \ArrowLine(53,60)(73,100)
        \Vertex(20,20){1.5} \Vertex(37,80){1.5}
        \GlueArc(150,20)(130,152,180){4}{9}
      }}
      \put(120,45){$B^0$}
      \put(195,-10){$\eta^{(\prime)}$}
      \put(195,105){$\bar D^{(*)0}$}
      \put(180,38){\small{}}
      \put(180,55){\small{}}
      \put(140,75){\small{$\bar{b}$}}
      \put(195,-30){(d)}
   \end{picture}
 }}
 \caption{Annihilation Diagrams contributing to the $B^0\to \bar D^{(*)0} \eta^{(\prime)} $
 decays.}
\end{figure}

 The $B^0\to \bar D^{(*)0} \eta^{(\prime)} $ decay rate has the expression as,
 \begin{equation}
 \Gamma =\frac{1}{128\pi}G_F^2|V_{cb}|^2|V_{ud}|^2m_B^3
 |{\cal M}|^2\;.
 \label{dr}
 \end{equation}
  Including the hard part and the meson wave functions,
   the $B^0\to \bar D^{(*)0}\eta^{(\prime)}$ decay amplitude is written as
 \begin{eqnarray}
  {\cal M} (B^0 \to \bar D^{(*)0} \eta^{(\prime)} )
  &=&  f_{D^(*)}\xi_{\rm int} + f_B\xi_{\rm exc}
       +{\cal M}_{\rm int} +{\cal M}_{\rm exc} \;,
 \label{M2}
 \end{eqnarray}
 where $f_B=190$ MeV, $f_{D}=f_{D^*}=240$ MeV  are the
 $B$  and $D^{(*)}$ meson decay constants, respectively.
  The
 functions   $\xi_{\rm int}$, and $\xi_{\rm exc}$
 denote the   internal $W$-emission, and
 $W$-exchange contributions, which come from Figs.~1(a) and 1(b),
 Figs.~2(a) and 2(b), respectively.
 The functions
 ${\cal M}_{\rm int}$, and ${\cal M}_{\rm exc}$ represent the
   internal $W$-emission, and
 $W$-exchange contributions, which come from Figs.~1(c) and 1(d),
 Figs.~2(c) and 2(d),   respectively.
 The expressions of the four functions are already shown in the
 appendix of ref.\cite{dpi} for $B\to D\pi$ decays. One need only
 replace the pion wave function   by the $\eta^{(\prime)}$ wave
 function in those expressions.

  In FA, only factorizable contribution of $\xi_{int}$ (Fig.1(a)(b)) has been
  considered.     Since $\xi_{int}$ is proportional to the small
  Wilson coefficient $a_2 = C_1+C_2/3$, the branching ratios
  predicted in FA is smaller than the experiments.
  Now in PQCD approach, all the topologies, including both factorizable and nonfactorizable
  ones, and also annihilation type ones have been taken into account.
  In fact the non-factorizable contribution $M_{int}$, which is proportional
  to the large Wilson coefficient $C_2/3$  is the dominant
  contribution in the $B^0 \to \bar D^{(*)0} \eta^{(\prime)} $   decays.
  The reason is that the two non-factorizable diagrams in Fig.1(c)
  and (d) do not cancel each other like the $B$ to two light meson
  decays, where the distribution amplitudes of wave function are
  symmetric \cite{KLS,LUY}. The large difference of $\bar c$ and
  $u$ quark mass makes the contribution of $M_{int}$ large.
Very recently, the soft collinear effective theory also confirms
that the non-factorizable $M_{int}$ dominate over the contribution
of $f_{D^{(*)}} \xi_{int}$ \cite{sc}.

\section{NUMERICAL RESULTS}
           \label{s3}

  As stated in the above section, we need various wave functions in
  our numerical calculations.    Considering the previously
  calculations of other decay channels \cite{dpi,TLS,KLS,LUY,kou},
  the $B$ meson wave function has been determined as
  \begin{eqnarray}
  \phi_B(x,b)=N_Bx^2(1-x)^2
  \exp\left[-\frac{1}{2}\left(\frac{xM_B}{\omega_B}\right)^2
  -\frac{\omega_B^2 b^2}{2}\right]\;,
  \label{os}
  \end{eqnarray}
  where the shape parameter is chosen as $\omega_B=0.4$
  GeV. The normalization constant $N_B$ is related to the decay constant
  $f_B$ through
  \begin{eqnarray}
  \int dx\phi_B(x,0)=\frac{f_B}{2\sqrt{6}}\;.
  \end{eqnarray}
  The $D^{(*)}$ meson distribution amplitude is given by
  \begin{eqnarray}
  \phi_{D^{(*)}}(x)=\frac{3}{\sqrt{6}}f_{D^{(*)}}
  x(1-x)[1+C_{D^{(*)}}(1-2x)]\;,
  \label{phid}
  \end{eqnarray}
  with the shape parameter $C_D=C_{D^*}=0.8\pm 0.2$ \cite{dpi}. The range of $C_{D^{(*)}}$ was
  extracted from the $B\to D^{(*)} l\bar \nu$ decay spectrum at large recoil
  assuming $\omega_B=0.4$ GeV for the $B$ meson wave function \cite{TLS}.
  We do not consider the variation
  of $\phi_{D^{(*)}}$ with the impact parameter $b$, since the current
  data are not yet sufficient to control this dependence.
 The light  $\eta^{(\prime)}$ meson wave functions are chosen to be the same
 as the pion wave function according to isospin symmetry,
 since the relevant  valence quarks here are mainly $u\bar u$ and $d\bar d$.

\begin{table}
\begin{center} \caption{PQCD predictions with one angle mixing formalism (I) and
two angle mixing formalism (II) and experimental data (in units of
$10^{-4}$)
               of the $B^0\to \bar D^{(*)0} \eta^{(\prime)} $ branching ratios.}
\begin{tabular}{|l| l| c| c|c|c|c|}
\hline {Decay mode}             &     PQCD (I)  & PQCD (II) &
Belle & BABAR &PDG
 \\ \hline
 $B^0\to \bar D^{0}\eta^{\prime}$ &  $1.7\sim 2.3$  &  $2.2\sim 2.6$  &  - &- &  $ <9.4 $\\
 ${ B}^0\to \bar D^{0}\eta $ &  $2.4\sim 3.0$ &  $2.6\sim 3.2$ & $1.4^{+0.6}_{-0.5}$
 &$2.41 \pm 0.50 $ & \\
 ${ B}^0\to \bar D^{0}\pi^0$ & $2.3\pm 0.1$ &   &  $ 3.1\pm 0.6$
 & $2.89\pm 0.48$ &\\
    ${ B}^0\to \bar D^{*0}\eta'$ & $2.0 \sim 2.7 $ &  $2.6\sim 3.2$&  & - &  $ < 14 $\\
 ${ B}^0\to \bar D^{*0}\eta $ &  $2.8\sim 3.5$ &  $3.1\sim 3.8$& $2.0^{+1.0}_{-0.9}$ &- &\\
 ${ B}^0\to \bar D^{*0}\pi^0$ & $2.8\pm 0.1$ &   &  $2.7 \pm 0.9 $& -& \\
     \hline
\end{tabular}
\label{pqcd}
\end{center}
\end{table}

In the numerical analysis we adopt
$$
 \Lambda_{\overline{\mathrm{MS}}}^{(f=4)} = 250\; {\rm MeV},~~
M_B = 5.2792\; {\rm GeV}, ~~M_W = 80.41\;{\rm GeV}.
$$
Choosing $|V_{cb}|=0.043$ and $|V_{ud}|=0.974$, we obtain the PQCD
predictions for the $B^0\to \bar D^{(*)0} \eta^{(\prime)} $
branching ratios shown in Table~\ref{pqcd}. For comparison, we
also list the $B^0 \to \bar D^{(*)0} \pi^0$ decay branching ratios
in this table.
 The theoretical uncertainty
comes   from the variation of the shape parameter for the
$D^{(*)0}$ meson distribution amplitude, $0.6 < C_{D^{(*)}} < 1.0$
and the $\eta$ and $\eta'$  mixing parameter $\theta =-17^\circ
\sim -11^\circ$ for the one-angle mixing formalism (I). The range
of parameters of two-angle formalism (II) are shown in
eq.(\ref{ran}). From numerical study, we notice that the branching
ratios do not
 vary much upon the variation of $C_{D^{(*)}}$. This can be seen
 from the numbers of $B^0 \to \bar D^{(*)0} \pi^0$ in Table~\ref{pqcd},
 since it only depends on this parameter.
 Most of the uncertainty of $B^0\to \bar D^{(*)0}\eta^{(\prime)}$
 decays is from the mixing parameter $\theta$ or $\theta_8$ and $\theta_0$.
The branching ratios of $B^0\to \bar D^{(*)0}\eta^{\prime}$
increase while $B^0\to \bar D^{(*)0}\eta$ decrease as the angle
$\theta$ getting larger.
 Other input parameters, such as parameters of $B$ meson  wave function
 also affect the branching ratios, but they are mostly constrained
 by other well measured decay channels, like $B\to \pi\pi$ \cite{LUY} and
 $B\to K\pi$ \cite{KLS} decays, etc.
 The uncertainties of PQCD approach itself mainly come from the
 unknown
 higher twist contributions and higher order calculations of
 $\alpha_s$ corrections. No numerical estimation of higher twist
 contribution exist, although it is expected to be suppressed. The
 higher order QCD calculation in $B\to \phi K$ decay shows that
 next-to leading order $\alpha_s$ correction may not be small in certain channels \cite{phik}.

  The recently observed class-II decay
  $B^0\to \bar D^{(*)0} \eta $ branching ratios
  are also listed in Table~\ref{pqcd} \cite{BelleC,CLEOC,PDG}.
 It is easy to see that, our results agree with the experimental
 measurements within errors.      Since we do not consider the extra gluon fusion contribution
 to    $B^0 \to \bar D^{(*)0}\eta'$ decay,  the not yet measured $B^0 \to \bar D^{(*)0}\eta'$ branching
 ratios  are a little  smaller than the  $B^0 \to \bar D^{(*)0}\eta$ branching
 ratios. But they are still comparable with the $B^0 \to \bar D^{(*)0}\pi^0$ branching
 ratios.
   The reason for this comparable result is that we apply  the assumption of exact
   isospin symmetry. We use the same wave function for $\eta^{(\prime)}$ and
   $\pi$ meson where the only difference is the decay constant.
    The difference of
   dynamics is the constructive or destructive contribution from
   annihilation  type diagrams. This destructive contribution
   makes the   $B^0 \to \bar D^{(*)0}\pi^0$ branching ratios
   smaller than that of    $B^0 \to \bar D^{(*)0}\eta$ decays.
   The numerical results also  show that the dominant contribution
   comes from the non-factorizable contribution $M_{int}$. The
   factorizable contribution $f_D\xi_{int}$ and annihilation
   contribution $M_{exc}$ are only 20-30\% of $M_{int}$.
   Factorizable annihilation contribution $f_B \xi_{exc}$ is
   negligible.

 \section{SUMMARY}      \label{s4}

 In this work, we calculate the branching ratios  of   $B^0\to \bar D^{(*)0} \eta^{(\prime)} $
 decays in the perturbative QCD approach with $k_T$ factorization, which is free of
 endpoint singularity.
 Belonging to class II decays in FA, these decays receive dominant contributions from the
 non-factorizable diagrams.     Naive factorization breaks down in
 these color suppressed modes.
 The branching ratios calculated in our approach for $B^0 \to \bar D^{(*)0} \eta $ agree with
 current
 experiments. We predict the branching ratios of $B^0 \to \bar
 D^{(*)0} \eta'$ without gluonic contributions are at the comparable size of $B^0 \to \bar D^{(*)0}
 \pi^0$, but smaller than  $B^0 \to \bar D^{(*)0} \eta $.
  They may be measured soon in the B factories.

\section*{Acknowledgments}

We thank E. Kou, H.n. Li and A.I. Sanda for discussions. This work
was supported by National Science Foundation of China under Grants
No.~90103013 and 10135060.

\end{document}